\documentclass[12pt]{article}
\usepackage[margin=1in]{geometry}
\usepackage{amsmath,amssymb}
\usepackage{graphicx}
\usepackage{hyperref}
\usepackage{placeins}
\newcommand{\mhyphen}{-}

\title{\textbf{A Comprehensive Mathematical and System-Level Analysis of Autonomous Vehicle Timelines}\\
\large{Integrating Complexity Theory, Reliability Growth, and ODD Modeling}}
\author{Paul Perrone -- \textit{Founder/CEO, Perrone Robotics, Inc.}}
\date{December 31, 2024}

\begin{document}

\maketitle

\begin{abstract}
Fully autonomous vehicles (AVs) continue to spark immense global interest, yet predictions on when they will operate safely and broadly remain heavily debated. This paper synthesizes two distinct research traditions---\textbf{computational complexity and algorithmic constraints} versus \textbf{reliability growth modeling and real-world testing}---to form an integrated, quantitative timeline for future AV deployment. We propose a mathematical framework that unifies NP-hard multi-agent path planning analyses, high-performance computing (HPC) projections, and extensive Crow-AMSAA reliability growth calculations, factoring in \textbf{operational design domain (ODD)} variations, severity, and partial vs.\ full domain restrictions. 

Through category-specific case studies (e.g., consumer automotive, robo-taxis, highway trucking, industrial and defense applications), we show how combining HPC limitations, safety demonstration requirements, production/regulatory hurdles, and parallel/serial test strategies can \emph{push out} the horizon for universal Level~5 deployment by up to several decades. Conversely, more constrained ODDs---like fenced industrial sites or specialized defense operations---may see autonomy reach commercial viability in the near-to-medium term. 

Our findings illustrate that while targeted domains can achieve automated service sooner, widespread driverless vehicles handling every environment remain far from realized. This paper thus offers a unique and rigorous perspective on \emph{why} AV timelines extend well beyond short-term optimism, underscoring how each dimension of complexity and reliability imposes its own multi-year delays. By quantifying these constraints and exploring potential accelerators (e.g., advanced AI hardware, infrastructure upgrades), we provide a structured baseline for researchers, policymakers, and industry stakeholders to more accurately map their expectations and investments in autonomous vehicle technology.
\end{abstract}

\section{Introduction}

Fully autonomous vehicle (AV) technology aims to eliminate human drivers entirely from the driving task (SAE Level 5). Despite extensive investments and limited early deployments, no comprehensive Level 5 system currently operates on \textit{all} roads and in \textit{all} weather conditions.

Over the years, many industry stakeholders have offered projections on when fully autonomous vehicles would become commonplace in everyday life. Some predictions have been overly optimistic and disproven year after year. Others have been more pessimistic, emphasizing the seemingly endless “edge cases” AVs must handle. Of course, many estimates are earnest attempts to forecast timing for business planning, research priorities, and strategic decision-making. Given the magnitude of opportunities, benefits, and global impact at stake, the need to continuously evaluate and refine these timelines remains pressing.

For more than two decades, I have endeavored to understand and project how pervasive AV deployment and commercialization might unfold. I have often relied on qualitative assessments—my own “blunt hammer” approach—to guide decisions in both business and personal contexts. Over time, however, I have grown increasingly curious about whether more objective, quantitative models could yield more accurate timeline projections.

Having followed numerous projections—often summarized in sound bites or short commentaries—and having made my own qualitative “gut-feel” forecasts, I became more conscious of the complexities and constraints that could push full AV adoption much further into the future than many realize. I know there are innumerable edge cases, that complexity differs dramatically across environments, that severe incidents can arise with large vehicles, and that untested vehicular platforms introduce additional hurdles. These factors have led me to suspect that some AV applications might be much farther off than we can easily imagine, given the complexity of dynamic and increasingly populated ground environments.

Hence, I began considering what quantitative methods might exist to predict the timeline for broad AV deployment more objectively. Could we develop mathematical approaches to produce better-informed projections—and might the latest AI tools help us locate and integrate a broader range of modeling ideas to improve accuracy?

This paper represents a first attempt to codify my evolving thoughts and assemble a system-level mathematical model for projecting AV timelines. Throughout the paper, I use “we” to acknowledge that the authorship includes not only my own efforts but also support from AI tools built on extensive human knowledge.

Two primary research traditions frame AV timeline estimation:

\begin{enumerate}
    \item \textbf{Computational Complexity \& Algorithmic Perspectives:}
    \begin{itemize}
        \item These highlight the vast state space of traffic environments, the NP-hard nature of planning with multiple dynamic agents under real-time constraints, and the unpredictability of real-world driving.
        \item They emphasize that even with advanced AI and Moore’s Law, the gap between theoretical performance and safe real-world operation may be larger—and slower to close—than often assumed.
    \end{itemize}
    \item \textbf{Reliability Growth Modeling \& Real-World Testing:}
    \begin{itemize}
        \item This tradition focuses on statistically demonstrating safety: achieving critical-failure rates (e.g., fatalities) on par with or better than human drivers.
        \item It requires billions or trillions of test miles to reduce failures to an acceptable threshold, as well as multiple product cycles and regulatory approvals.
    \end{itemize}
\end{enumerate}

The novelty here lies in combining these two perspectives into a cohesive mathematical framework for more robust timeline projections. We then refine these models further. After introducing the integrated models in Sections 2 and 3, we incorporate additional considerations and constraints in Sections 4 and 5 to arrive at a model for realistic deployment timelines. Section 6 provides an overview of the current deployment landscape, while Section 7 applies the integrated model to various AV categories. Section 8 presents a high-level summary of the findings, and Section 9 concludes with final remarks and avenues for future work.

\section{Complexity-Driven Analysis}
\label{sec:complexityAnalysis}

In this section, we explore how the sheer breadth of possible driving scenarios, combined with the exponential growth of multi-agent decision-making, creates substantial computational obstacles for fully autonomous vehicles. We begin by quantifying the state-space explosion inherent in traffic environments, where each agent’s position, velocity, and behavioral parameters multiply rapidly. Then, we discuss the NP-hard nature of multi-agent path finding (MAPF) and how real-time constraints intensify the associated demands for high-performance computing (HPC). Finally, we highlight practical heuristics that can partially tame this complexity, acknowledging that even with sophisticated optimization strategies, meeting the real-time requirements of SAE Level~5 autonomy may necessitate multi-decade advances in hardware and software maturity.

\subsection{State-Space Explosion}
\label{sec:stateSpace}
How do we quantify the complexity of the environment in which an AV operates? One approach is to look at the state space as a function of objects, their parameters, and the levels of variation among these parameters. 

\textbf{Number of Dynamic Objects $(n)$:}  
In a dense urban environment, an AV’s sensors (cameras, LiDAR, radar, etc.) might simultaneously observe and track multiple \emph{dynamic objects}:
\begin{itemize}
    \item Vehicles (cars, trucks, buses)
    \item Pedestrians
    \item Cyclists
    \item Animals or other moving obstacles
\end{itemize}
A typical conservative estimate is that around 50 of these dynamic objects could be relevant at any one time ($n = 50$). This represents a heavily trafficked intersection or congested street. Each object is \emph{dynamic} because it changes position, speed, direction, or behavior over time.

\textbf{Possible States per Object $(k)$:}  
To characterize a single object’s situation (position, velocity, orientation, behavior), one can discretize the parameter space. For example:
\begin{itemize}
    \item Let $d = 10$ be the number of parameters per object (position, velocity, acceleration, orientation, plus a few discrete behavior/intent states).
    \item Let $m = 100$ be the number of discrete levels for each parameter (e.g., dividing an axis into 100 increments).
\end{itemize}
Hence, for a single object:
\[
k = m^d = 100^{10} = 10^{20}.
\]
This $k$ counts all possible \emph{discretized} states for one moving agent, given $d$ parameters at $m$ levels each.

\textbf{Total State Space $(S)$:}  
When $n$ such objects each occupy one of $k$ states, the \emph{joint} or \emph{combined} state space is:
\[
S = k^n = \bigl(10^{20}\bigr)^{50} = 10^{1000}.
\]
Even for 50 objects, each with 10 parameters, we reach $10^{1000}$ total configurations—making exhaustive or brute-force planning intractable.

\subsection{Naive Multi-Agent Path Finding}
\label{sec:MAPFConstraints}

Many decision-making problems in autonomous driving require \emph{planning} and \emph{routing} for multiple moving agents simultaneously. A key formulation is \textbf{Multi-Agent Path Finding (MAPF)}, which involves computing feasible trajectories for multiple agents without collisions. MAPF is known to be NP-hard, meaning computational complexity grows exponentially with the number of agents:
\[
T_c = O\bigl(2^n\bigr),
\]
where \(n\) is the number of dynamic objects (vehicles, pedestrians, cyclists, etc.) that must be simultaneously considered.  

\paragraph{Naive Theoretical Computation.}
For instance, if one has \(n = 50\) relevant agents, a naive worst-case model might estimate 
\[
T_c \approx 2^{50} \approx 10^{15} 
\]
operations \emph{per planning cycle}.  If the AV system has an industry-standard reaction time of \(\sim100\text{\,ms}\), it requires:
\[
C_d \;=\; \frac{10^{15}\text{ operations}}{0.1\text{ s}} 
         \;=\; 10^{16}\text{ ops/s}.
\]
Similarly, for \(n = 60\), one might extrapolate 
\[
T_c \approx 2^{60} \approx 10^{18}\text{ operations per cycle},
\]
leading to 
\[
C_d = \frac{10^{18}}{0.1} = 10^{19}\text{ ops/s}.
\]
In a purely theoretical sense, raising \(n\) from 50 to 60 can inflate the required compute demand from \(10^{16}\) to \(10^{19}\) ops/s.

\subsection{Practical High Performance Computing (HPC) Demand}
\label{sec:PracticalHPCandChi}

While the naive \(\,2^n\,\) MAPF analysis gives an \emph{upper bound} on computational needs, real AV architectures rarely attempt a full-blown, exact multi-agent search across all dynamic objects every 100\,ms.  Instead, they employ a host of engineering heuristics that drastically cut the “effective” HPC requirement.  We capture these heuristics via an overall factor \(\chi\) (chi), yielding:
\[
C_d' \;=\; C_d \;\times\; \chi,
\]
where \(C_d\) is the naive HPC estimate (e.g.\ \(10^{19}\) ops/s for \(n=60\)), and \(\chi\in(0,\,1)\) is a reduction factor reflecting real-world considerations. 

\paragraph{Enumerating Real-World Reductions.} 
Rather than picking \(\chi\) arbitrarily, we can view it as a \emph{product} of multiple “\(p\) factors” (or \(\rho\) factors) that each reduce HPC load:

\[
\chi_{\mathrm{eff}} 
\;=\; 
p_1 \;\times\; p_2 \;\times\;\dots \;\times\; p_k,
\]
where each \(p_i\) (or \(\rho_i\)) represents a known reduction mechanism:
\begin{itemize}
  \item \textbf{Limiting active objects}: Only a subset (e.g.\ 20 of the 60) truly matter for immediate collision checks \(\implies p_1\approx 0.33\), assume ranges 0.2-0.5.
  \item \textbf{Temporal slicing}: Comprehensive MAPF may run at 500\,ms intervals (not 100\,ms), or partial updates at 100\,ms, etc. \(\implies p_2\approx 0.2\), assume ranges 0.1-0.3.
  \item \textbf{Coarse modeling of distant agents}: Another factor \(p_3\approx 0.2\), assume ranges 0.1-0.3.
  \item \textbf{Local planning vs.\ global}: Some fraction of objects can be “fenced off” via local-lane heuristics \(\implies p_4\approx0.1\), assume ranges 0.1-0.3.
  \item \textbf{ODD restrictions}: e.g.\ geofenced routes, lower speed \(\implies p_5\approx0.5\), assume ranges 0.1-1.0.
\end{itemize}

By multiplying these \(p_i\), one obtains an explicit \(\chi_{\mathrm{eff}}\).  For example, if each factor is around 0.2–0.3, the product may easily reach 0.0001.  Thus, the “\(\sim10^{19}\,\mathrm{ops/s}\)” from naive MAPF gets reduced to around \(10^{16}\,\mathrm{ops/s}\). 

\paragraph{Stage-Based HPC Targets.}
We then interpret \(\chi\) (or \(\chi_{\mathrm{eff}}\)) differently for earlier vs.\ more advanced stages of autonomy:
\begin{itemize}
  \item \textbf{Limited ODD - Moderate Heuristics Stage:} 
        \(\chi_{\mathrm{eff}}\approx0.01\) to 0.0001.
  \item \textbf{Full L5 - Extensive Heuristics Stage:}
        \(\chi_{\mathrm{eff}}\approx0.1\) or 0.01, if we assume strong domain-pruning but must handle more universal roads.
\end{itemize}
Hence, if the naive MAPF yields \(C_d=10^{19}\,\mathrm{ops/s}\) for \(n=60\), then for a Stage~2 environment with extensive partial constraints, we might adopt \(\chi_{\mathrm{eff}}=0.001\).  That leads to an effective HPC demand 
\[
 C_d' 
   = 
   10^{19}\,\mathrm{ops/s}\,\times\,0.001
   =
   10^{16}\,\mathrm{ops/s}.
\]
For an earlier stage with even more geofencing, one might push \(\chi_{\mathrm{eff}}\to0.0005\).  Conversely, a less pruned ODD might yield \(\chi\approx 0.01\).  

\subsection{Timeline to Achieve HPC}
\label{sec:techReadiness}

Hardware and software maturity do not appear overnight. We have to understand how quickly high-performance computing (HPC) and related automotive-grade hardware/software solutions approach maturity. Historically, consumer-focused electronics often benefited from rapid transistor scaling (``Moore's Law'') every 18--24\,months, but automotive deployments face additional constraints: strict real-time requirements, lengthy qualification cycles, and high reliability/safety standards that slow the \emph{effective} pace of technology readiness. 

Even if a form of Moore’s Law continues, bridging the order-of-magnitude gaps in real-time decision-making can still require \emph{one to three decades}, depending on the \emph{doubling rate} and on additional complexities for safety-critical design. As of 2024, high-end data-center GPUs can achieve around $10^{15}$ ops/s, whereas in-vehicle compute is often an order or two lower ($10^{13}$ ops/s) due to size, power, and thermal constraints. Yet from Section~\ref{sec:stateSpace} and Section~2.2 (Multi-Agent Path Finding), we see that naive or near-exact planning might demand roughly $10^{16}$ ops/s.  

\paragraph{HPC Growth Model:} A simple HPC growth model might assume a regular doubling every $T_d$ years (e.g., every 2.5 years). Historically, Moore’s Law is often cited as doubling transistor density every 18–24 months (e.g. $T_d$ every 1.5-2 years), but transistor scaling has since slowed, packaging costs have risen, and specialized accelerators (e.g. GPUs, AI TPUs) do not always follow CPU-centric timelines. In safety-critical automotive contexts, certification, real-time constraints, and operational robustness (e.g. temperature, vibration, longer lifecycles) further lag effective performance gains relative to raw transistor advances. Even if data-center GPUs reach $10^{16}$ ops/s, powering and cooling them at scale in cars introduces additional engineering delays. Consequently, choosing $T_d$ = 2.5 years here reflects a moderate HPC growth estimate for automotive usage as of late 2024.

Thus, if $C_c$ is the current on-vehicle compute performance (around $10^{13}$ ops/s in 2024), then:

\[
C(t) \;=\; C_c \times 2^{\frac{t}{T_d}},
\]

where $t$ is measured in years, $T_d \approx 2.5$. For instance, if the target demand $C_d \approx 10^{16}$ ops/s, we can estimate the time to close that three-order-of-magnitude gap:

\[
\frac{C_d}{C_c} 
\;=\; 
\frac{10^{16}}{10^{13}} 
\;=\; 10^3 
\;\;\Rightarrow\;\; 
2^{\frac{t}{2.5}} = 10^3.
\]

Taking $\log_2$ of both sides yields:

\[
\frac{t}{2.5} = \log_2(10^3) \,\approx\, \log_2(1000) \,\approx\, 9.97,
\]

hence $t \approx 2.5 \times 9.97 \approx 25$ years. This aligns with the one-to-three decade estimate, although actual progress can be faster or slower if HPC innovation accelerates or plateaus. Moreover, for \emph{safety-critical} systems, simply hitting the raw ops/s target might be insufficient; additional redundancy, real-time constraints, and certification overhead extend the effective timeline.

In other words, we  must account not just for improvements in raw performance, but also for the engineering, safety, and validation challenges. A plateau in HPC could slow full readiness. Conversely, breakthroughs (quantum, optical, specialized AI accelerators) could dramatically boost the rate of improvement.

\paragraph{Years to Achieve Compute Demand:}
Thus, the HPC growth model underscores how bridging each successive order of magnitude in compute can require multi-year increments. This dynamic becomes particularly critical once we incorporate real-time demands, NP-hard path planning, and the need for robust sensor fusion in safety-critical AV domains.

In later sections, we refer to this \emph{HPC feasibility horizon} $\boldsymbol{t}$ as
\[
T_{\mathrm{comp}} \;=\; T_{\mathrm{d}} \log_2(\frac{C_d \times \chi}{C_c})
\]

That is, $T_{\mathrm{comp}}$ indicates the number of years from the current (2024) baseline until the required compute performance is reached (or at least approached) for a given AV domain. While breakthroughs or slowdowns could shift this timeline, it underscores how bridging successive orders of magnitude in compute often spans a decade or more—particularly under safety-critical constraints.

Note that we choose 2024 as the baseline year because it aligns with current HPC and pilot deployment data at the time of this analysis. This does not necessarily assume AV development is starting from zero in 2024. Rather, it just means that our HPC and deployment references are pinned to late-2024 capabilities. Many programs have prior software, but the hardware environment (e.g.\ $10^{13}$ ops/s in-vehicle) is typical of 2024 commercial solutions.

\paragraph{HPC vs.\ Technology Readiness:}
Achieving a target HPC demand does not automatically render an AV system ready for deployment. Rather, the HPC timeline addresses only the compute aspect of technology readiness, while other factors—such as sensor maturity, software development, and supporting infrastructure—also play pivotal roles. In this facet of our model, however, we assume HPC growth effectively bounds overall technology readiness. We further assume that (i) software will be available in parallel to exploit the growing HPC capabilities, (ii) sensor technology is already near a readiness level suitable for autonomous operations, and (iii) required infrastructure will not become a gating factor relative to HPC growth.

\subsection{Additional Algorithmic and Environmental Factors}

Beyond the raw state-space and MAPF challenges, further complications affect real-world AV deployment:

\begin{itemize}
    \item \textbf{Machine Learning Generalization:} Deep neural networks still struggle to handle all corner cases, facing non-trivial generalization errors for real-world driving complexity.
    
    \item \textbf{Undecidability and Rice’s Theorem:} All non-trivial, semantic properties of programs are undecidable, meaning no algorithm can guarantee correct behavior in \emph{all} scenarios.
    
    \item \textbf{Adversarial Examples:} Neural nets are vulnerable to small perturbations in sensor data that can produce large errors in output. Designing robust systems for the vast input space remains extremely challenging.
    
    \item \textbf{Stochastic Nature:} Weather, road conditions, and unpredictable human behaviors introduce randomness. While Markov Decision Processes (MDPs) can model uncertainty, large MDPs become computationally infeasible.
    
    \item \textbf{Chaotic Dynamics:} Minor estimation errors can significantly alter trajectory outcomes (sensitive dependence on initial conditions).
\end{itemize}

These issues compound the complexity of real-time decision-making and underscore why a purely incremental engineering approach may not suffice.

\section{Reliability Growth Modeling}
\label{sec:ReliabilityGrowth}

While computational complexity highlights the scale of the driving problem, demonstrating that an AV can safely operate within that high-dimensional state space requires rigorous reliability assessments. In this section, we introduce statistical reliability-growth frameworks that quantify how quickly critical failures decrease as vehicles accumulate real or simulated test miles. Specifically, we present the Crow-AMSAA (power-law) model, a well-established method in automotive and aerospace programs, and compare it with Poisson-based approaches to determining mileage requirements for safety-validation goals. By integrating these reliability metrics with the constraints discussed in Section~\ref{sec:complexityAnalysis}, we form a more holistic picture of the timeline needed to achieve robust, failure-resistant autonomous systems.

\subsection{Crow-AMSAA (Power-Law) Approach}
\label{sec:CrowAMSAA}

The \textbf{Crow-AMSAA} (also known as the \emph{Crow-Armstrong-AMSAA}) model is a power-law method widely used in reliability growth analysis, originally developed for military applications but also applied to automotive and aerospace. It characterizes how the failure rate of a system decreases as more operational (or test) time accumulates, assuming ongoing improvements or “fixes” after each failure.

Formally, let:
\[
\lambda(t) \;=\; \alpha \, t^{-\beta} \,\times\, s,
\]
where:
\begin{itemize}
    \item \(t\) is the total accumulated \emph{test miles} (or hours),
    \item \(\beta\) is the \emph{reliability growth exponent}, indicating how quickly the failure rate decreases with more test time,
    \item \(\alpha\) is an empirically fitted constant from early test data,
    \item \(s\) is a \emph{severity factor}, representing the average number of fatalities \emph{when a failure incident does result in fatalities}.
\end{itemize}

This is a tailored equation adapted for our needs here in this paper to express a target failure rate as a function of reliability growth, a severity factor, test miles, and a constant. To meet a target failure rate \(\lambda_{\mathrm{target}}\) (e.g., aiming for one fatality per \(10^8\) or \(10^9\) miles), we want:
\[
\lambda\bigl(R(t)\bigr)
\;=\; 
\alpha \,\bigl(R(t)\bigr)^{-\beta}\,\times\, s
\;\;\le\;\; \lambda_{\mathrm{target}}.
\]
Solving for \(\mathbf{R(t)}\):
\[
\alpha\,s\,\bigl(R(t)\bigr)^{-\beta}
\;\;\le\;\;
\lambda_{\mathrm{target}},
\]
\[
\bigl(R(t)\bigr)^{-\beta}
\;\;\le\;\;
\frac{\lambda_{\mathrm{target}}}{\alpha\,s},
\]
\[
R(t)^\beta
\;\;\ge\;\;
\frac{\alpha\,s}{\lambda_{\mathrm{target}}},
\]
\[
R(t)
\;\;\ge\;\;
\Bigl(\frac{\alpha\,s}{\lambda_{\mathrm{target}}}\Bigr)^{\!\frac{1}{\beta}}.
\]
Hence, the \textbf{mileage requirement} \(R(t)\) (in miles) to achieve or surpass the target failure rate \(\lambda_{\mathrm{target}}\) under Crow-AMSAA depends on the exponent \(\beta\), the empirical constant \(\alpha\), and the severity factor \(s\). 

\paragraph{Illustrative Example.}
Suppose:
\[
\beta \approx 0.3 \text{–} 0.5, 
\quad
s = 2 
\;(\text{e.g.\ on average 2 fatalities occur when a failure incident does result in fatalities}),
\quad
\alpha \approx 10^{-2},
\quad
\lambda_{\mathrm{target}} = 10^{-8}.
\]
Then
\[
R(t)
\;\;\ge\;\;
\Bigl(\frac{10^{-2} \times 2}{10^{-8}}\Bigr)^{\!\frac{1}{\beta}}
\;=\;
\Bigl(\,2\times10^{6}\Bigr)^{\!\frac{1}{\beta}}.
\]
- If \(\beta=0.3\), \(R(t)\approx (2\times10^6)^{3.33}\approx 10^{21}\) miles (very large).  
- If \(\beta=0.5\), \(R(t)\approx (2\times10^6)^{2}=4\times10^{12}\) miles (still large, but less so).

These wide ranges illustrate why the \emph{rate of reliability growth} (\(\beta\)) and the \emph{severity factor} (\(s\)) strongly affect how many miles are needed before meeting a desired failure threshold. In real AV programs, \(\beta\) can rise over time (faster learning/improvements), reducing the exponent’s impact on mileage requirements.

\subsubsection{A Note on the Empirical Constant 
(\texorpdfstring{$\alpha$}{alpha})}

In the Crow-AMSAA (power-law) reliability growth model, \(\alpha\) appears as part of the failure-rate expression.  Here, \(\alpha\) can be viewed as an “initial” or “baseline” failure‐rate coefficient:

\begin{enumerate}
    \item \textbf{Conceptual Role:}
    \begin{itemize}
        \item A larger \(\alpha\) implies that, before learning and improvements significantly kick in, the initial or early-phase failure rate is higher. Conversely, a very small \(\alpha\) suggests a lower initial baseline failure tendency.
        \item As reliability engineers gather real test data, they typically fit \(\alpha\) along with \(\beta\) to reflect how quickly failures are discovered and fixed.
    \end{itemize}
    \item \textbf{Example Values in Practice:}
    \begin{itemize}
        \item \(\alpha\) = 0.01 might be plausible if the AV program expects a relatively moderate initial failure rate once serious on-road testing begins.
        \item \(\alpha\) = 0.0001 can be used to represent an even lower initial baseline failure rate. This might be relevant if the AV has already been iterating heavily, or if significant portions of the system are well-matured from prior development.
    \end{itemize}
    \item \textbf{Which Is “Correct”?}
    \begin{itemize}
        \item Technically, \(\alpha\) is not universal but must be calibrated based on how advanced the AV system is when applying the Crow-AMSAA method. Different approaches will yield an \(\alpha\) that matches observed or assumed failure rates in the early test phase.
        \item In the example above, \(\alpha\) = 0.01 simply demonstrates how big R(t) can become if \(\alpha\) is moderately sized and \(\beta\) is small. In some subsequent analyses in this paper, we might assume a smaller \(\alpha\) because we are illustrating a scenario where the AV has already had multiple cycles of refinement (thus a lower “starting” fail rate).
    \end{itemize}
    \item \textbf{Acceptable Ranges and Stage-Specific Choices:}
    \begin{itemize}
        \item For early, pilot-phase prototypes—where the system is still fairly unproven—\(\alpha\) values around 0.01 or even 0.1 might be used. This signals that many initial failures are expected until engineers iterate on them.
        \item For more mature or vehicles, or if substantial pre-development has already occurred (extensive simulation, partial autopilot experience), one might see \(\alpha\) in the 0.001 to 0.0001 ballpark if we assume a system has undergone major improvements before expanding their deployments further.
    \end{itemize}
\end{enumerate}

\subsection{Quantitative Safety Validation Requirements}
\label{sec:QuantSafetyVal}

In many safety validation approaches, we also want to ensure the AV system's failure rate \(\lambda\) remains below some \(\lambda_{\mathrm{target}}\) with a high confidence \(C\) (e.g., 95\% confidence). A common Poisson-based approach sets the \emph{required test mileage} \(R(t)\) by specifying:

\[
R(t) 
\;=\;
\frac{-\ln(1 - C) \;\times\; SF}{\lambda_{\mathrm{target}}},
\]

where:
\begin{itemize}
    \item \(C\) = desired confidence (e.g., 0.95),
    \item \(\lambda_{\mathrm{target}}\) = max acceptable failure rate (e.g., \(10^{-8}\) per mile),
    \item \(SF\) = a safety factor (e.g., 2.0),
    \item \(\mathbf{R(t)}\) = the \textbf{mileage requirement} (in miles) under the zero-failures assumption.
\end{itemize}

\paragraph{Origin of the Formula.}
A typical derivation uses a simplifying assumption that \textbf{no failures} occur during the test mileage \(R(t)\). For a Poisson process with rate \(\lambda\), the probability of observing zero failures in \(R(t)\) miles is:

\[
P(\text{0 failures}) 
\;=\; 
e^{-\lambda\,R(t)}.
\]

If we want to be at least \(C\) confident that \(\lambda \le \lambda_{\mathrm{target}}\), we require:

\[
P(\text{0 failures})
\;=\;
e^{-\lambda_{\mathrm{target}}\,R(t)\div SF}
\;\;\ge\;\;
C,
\]

where we introduced \(SF\) as an extra buffer for unknowns or potential underestimation. Rearranging to solve for \(R(t)\) yields:

\[
e^{-\lambda_{\mathrm{target}}\,R(t)\div SF}
\;\;\ge\;\;
C
\;\;\implies\;\;
-\lambda_{\mathrm{target}}\,R(t)\div SF
\;\;\ge\;\;
\ln(C),
\]

\[
R(t) 
\;\;\ge\;\; 
\frac{-\ln(C) \times SF}{\lambda_{\mathrm{target}}}.
\]

Since \(C\) is typically something like \(0.95\), we may rewrite \(\ln(C)\) as \(-\ln(1-C)\) for convenience, giving:

\[
R(t) 
\;=\;
\frac{-\ln(1 - C) \times SF}{\lambda_{\mathrm{target}}}.
\]

\paragraph{Illustration.}
Suppose we want to show a failure rate \(\lambda_{\mathrm{target}} = 7.1\times10^{-9}\) (which is about 50\% better than the human baseline of \(1.42\times 10^{-8}\)), with \(C=0.95\) and \(SF=2.0\). Then:

\[
R(t)
\;=\;
\frac{-\ln(1 - 0.95) \;\times\; 2.0}{7.1\times10^{-9} }
\;\approx\;
844\,\text{million miles}.
\]

Hence, under these assumptions, about 844 million miles of failure-free testing (or an equivalent demonstration of low failure rate) may be required just to confirm that the AV meets \(\lambda_{\mathrm{target}}\) at 95\% confidence with a safety factor of 2.

This relatively straightforward calculation can balloon for stricter confidence requirements, smaller \(\lambda_{\mathrm{target}}\), or higher safety factors, illustrating why many AV programs have turned to extensive real-world plus simulated miles. Meanwhile, the \emph{Crow-AMSAA} approach (discussed in Section~\ref{sec:CrowAMSAA}) adds a power-law learning curve, refining how \(\lambda\) evolves with more testing and iterative improvements.

\subsection{Reliability Demonstration (\texorpdfstring{$T_{\mathrm{rel}}$}{Trel})}

The \emph{Crow-AMSAA} reliability growth model and the \emph{Poisson-based} safety validation approach each yield a required mileage \(R(t)\) to demonstrate that an AV’s failure rate has reached an acceptable threshold. The required mileage modeling (with Crow-AMSAA) can incorporate:
\begin{itemize}
    \item \(\beta\): a reliability growth exponent, indicating how quickly failures decrease with more mileage.
    \item \(s\): a severity factor, reflecting how many fatalities occur \emph{when} an incident does result in fatalities.
\end{itemize}
Either way—whether from Crow-AMSAA or from the Poisson-based approach—the outcome is a required \(R(t)\) (possibly billions/trillions of miles). Once \(R(t)\) is known, we then determine how many \emph{calendar years} that might take, which defines:

\[
T_{\mathrm{rel}} \;=\; \frac{R(t)}{M},
\]

where \(M\) is the annual testing capacity (real + simulated miles). Large \(R(t)\) plus modest \(M\) can easily lead to multi-decade \(T_{\mathrm{rel}}\) horizons unless advanced reliability growth techniques (raising \(\beta\)) or domain restrictions (reducing \(s\) or focusing on simpler roads) drastically cut the required mileage.

\paragraph{Practical Considerations.} 
\begin{itemize}
    \item If an AV program conducts \(\mathbf{10^9}\) miles/year of real + simulated testing and needs \(\mathbf{10^{11}}\) total miles, that alone is 100 years unless either \(\beta\) speeds up or one uses additional corner-case coverage methods. 
    \item In contrast, if \(\beta\approx 0.5\) (faster reliability growth), or we reduce \(s\) by focusing on a simpler domain, \(R(t)\) may drop enough that \(T_{\mathrm{rel}}\) becomes feasible on a 5--10 year horizon.
\end{itemize}

Thus, \(\mathbf{T_{\mathrm{rel}}}\) is the \emph{time dimension} of reliability demonstration, bridging the theoretical mileage requirement \(R(t)\) to a real-world calendar schedule.

\subsection{ODD Complexity Factor (\texorpdfstring{$\gamma$}{gamma})}
\label{sec:ODD-complexity}

When evaluating reliability growth, the complexity of the deployment should be considered. Less complexity often means fewer or simpler components. Be it slower speeds that permit shorter range sensors, or a less complex environment that allows for simpler software implementations. This \emph{operational design domain} (ODD) itself can range from low-speed, fair-weather private roads to dense urban highways with unpredictable traffic. We quantify this via a dimensionless complexity factor:
\[
\gamma = \sum_i \bigl(w_i \times c_i\bigr),
\]
where $w_i$ are the weights for major complexity dimensions (weather, traffic density, road type, speed range), and each $c_i$ is a 0--1 score of how challenging that dimension is. For example:
\begin{itemize}
    \item Weather (weight 0.3): fair-only $\to c_{\mathrm{weather}}=0.2$ vs.\ all-weather $c_{\mathrm{weather}}=1.0$.
    \item Traffic density (weight 0.25): light $\to 0.2$ vs.\ heavy/unpredictable $\to 1.0$.
    \item Road complexity (weight 0.25): highways vs.\ complex urban grids.
    \item Speed range (weight 0.2): low speeds (0.2) vs.\ 0--80 mph (1.0).
\end{itemize}
Thus, smaller $\gamma$ indicates a simpler ODD (like a low-speed shuttle in fair weather), while large $\gamma$ ($>0.7$) indicates a full-scale urban environment. As $\gamma$ increases, more test miles, HPC resources, and advanced algorithms are required to handle the bigger effective state space. 

Note that $\gamma$ is described as a dimensionless number that reflects how challenging or broad the operational design domain is. If it is less than 1.0, it might indicate a simplified or restricted environment (e.g., low-speed roads, fair weather). If it is greater than 1.0, it might indicate a domain that is more complex than some baseline expectation (e.g., dense, chaotic urban environment). If it is exactly 1.0, it might be the “baseline” or nominal ODD we compare everything to.

\subsection{ODD Reduction Factor (\texorpdfstring{$\delta$}{delta})}
\label{sec:ODD-reduction}

We also want to capture a representation of how much the ODD is intentionally restricted compared to a full real-world scenario set. For this, we use a a dimensionless parameter, \emph{$\delta$}, representing a fraction of how much the ODD is intentionally restricted. For instance, if we only operate in fair weather and moderate speeds, that might represent $\sim30\%$ of typical traffic scenarios, so we set $\delta$ = $0.3$.

For clarity, the ODD complexity factor ({$\gamma$}) reflects the intrinsic breadth and difficulty of a vehicle’s operational design domain, accounting for such dimensions as weather extremes, traffic density, road geometry, and speed range. By contrast, the ODD reduction factor ({$\delta$}) (often a fraction less than 1) captures how much of that potential domain the system intentionally excludes to simplify deployment (e.g. only daytime or low-speed roads). In practice, {$\gamma$} is an inherent multiplier that increases or decreases the total required test miles based on how “busy” or “rich” the environment is, while {$\delta$} provides a further scaling by removing challenging scenarios outright. For instance, a system might have a moderately high complexity ({$\gamma$} greater than 1), yet still reduce its real-world trial burden by restricting operation to 30 percent of possible weather and traffic conditions—applying an ODD reduction factor {$\delta$} of 0.3.

\subsection{Refined Reliability Demonstration Time (\texorpdfstring{$T_{\mathrm{rel}}'$}{Trel'})}
\label{sec:TrelPrime}

In earlier sections, we introduced two primary methods for determining the \textbf{required mileage} $R(t)$ needed for an AV system to reach a desired failure rate $\lambda_{\mathrm{target}}$:

\begin{enumerate}
    \item \textbf{Crow-AMSAA (Power-Law) Model} (Section~\ref{sec:CrowAMSAA}), yielding
    \begin{equation}
    \label{eq:R_CrowAMSAA}
    R_{\!\mathrm{Crow}}(t) 
    \;=\; 
    \Bigl(\frac{\alpha\,s}{\lambda_{\mathrm{target}}}\Bigr)^{\!\frac{1}{\beta}},
    \end{equation}
    where \(\alpha\), \(\beta\), and \(s\) describe iterative reliability growth (see Section~\ref{sec:CrowAMSAA}).

    \item \textbf{Poisson/Zero-Failures Confidence Model} (Section~\ref{sec:QuantSafetyVal}), yielding
    \begin{equation}
    \label{eq:R_Poisson}
    R_{\!\mathrm{Poisson}}(t) 
    \;=\;
    \frac{-\ln\bigl(1 - C\bigr) \,\times\, SF}{\lambda_{\mathrm{target}} },
    \end{equation}
    where \(C\) is confidence (e.g.\ 0.95) and \(SF\) is a safety factor (e.g.\ 2.0).
\end{enumerate}

Either approach provides a way to compute $R(t)$ (the total test mileage required) based on different assumptions or parameters. In practice, an analysis effort might use one model or a hybrid. For the general discussion here, we can treat
\begin{equation}
\label{eq:GenericRofT}
R(t) 
\;=\; 
\text{(some function of }\beta,\,s,\,\alpha,\,C,\,SF,\ldots\bigl).
\end{equation}

\paragraph{From Required Miles $R(t)$ to a Calendar Timeline $T_{\mathrm{rel}}$.}
Once $R(t)$ is determined, we define:
\begin{equation}
\label{eq:Trel_simple}
T_{\mathrm{rel}}
\;=\;
\frac{R(t)}{M},
\end{equation}
where $M$ is the \emph{annual testing capacity} (the total real + simulated miles per year). This simple division converts “total required miles” into a “calendar time” to achieve those miles.

\paragraph{Incorporating ODD Considerations.}
We refine $T_{\mathrm{rel}}$ further by noting that (1) \textbf{ODD complexity} $(\gamma)$ typically \emph{increases} the effective mileage needed, and (2) an \textbf{ODD reduction factor} $(\delta)$ can lower the total scenario set if one restricts operations. Hence, we define a \emph{refined} time to deployment:

\begin{equation}
\label{eq:TrelPrime}
T_{\mathrm{rel}}'
\;=\;
\frac{\,R(t)\,\times\,\gamma\,\times\,\delta\,}%
     {\,\text{M}\,\\}.
\end{equation}

\noindent
\textbf{Explanatory Steps:}
\begin{enumerate}
    \item \emph{Scale for ODD complexity:} Multiply $R(t)$ by $\gamma$ to represent a more challenging environment requiring additional test coverage.
    \item \emph{Apply ODD limitation:} Further divide by an ODD reduction factor $\delta$ (e.g.\ 0.3) if restricting speed, weather, or domain to a fraction of possible scenarios.
    \item \emph{Convert to years:} Divide by testing capacity \texttt{M}, which is the total (real + simulated) miles per year.
\end{enumerate}

\section{Combined Model}
\label{sec:combinedModel}

Having examined the challenges posed by computational complexity (Section~\ref{sec:complexityAnalysis}) and the requirements for rigorous safety validation (Section~\ref{sec:ReliabilityGrowth}), we now merge these perspectives into a single, unified framework for projecting autonomous-vehicle (AV) timelines. This “combined model” accounts for both the hardware and software demands of real-time decision-making and the mileage-based reliability demonstrations needed to meet critical failure-rate thresholds. Additionally, it incorporates production cycles, regulatory approvals, and operational design domain (ODD) factors, all of which can impose multi-year constraints on any real-world deployment of Level~5 autonomy.

\subsection{Product Engineering and Regulatory Timelines (\texorpdfstring{$T_{\mathrm{prod+reg}}$}{Tprod+reg})}
\label{sec:prodReg}

Meeting complexity demands (from Section~\ref{sec:complexityAnalysis}) and reliability targets (from Section~\ref{sec:ReliabilityGrowth}) are only two dimensions of deploying autonomous vehicles. In parallel, \textbf{product engineering and regulatory approval} often require substantial lead times before a new design can be manufactured and sold at scale. We capture this in a term \(\mathbf{T_{\mathrm{prod+reg}}}\), which reflects:

\begin{itemize}
    \item \textbf{Engineering \& Production Integration:}
    \begin{itemize}
        \item Typical automotive development cycles for a new or significantly modified vehicle platform can run 3--5 years.
        \item Even incremental changes (e.g., adding more sensors or updated control units) may need 1--2 years of redesign and testing before entering mass production.
        \item Novel or unique vehicle platforms may also require additional engineering and production time versus traditional platforms for additional design and testing cycles to eek out new configurations for safety assurance.
    \end{itemize}

    \item \textbf{Regulatory Approvals:}
    \begin{itemize}
        \item Government agencies and safety regulators as well as safety standards (e.g., NHTSA or FMVSS in the U.S.) must certify that a vehicle meets crashworthiness standards and other requirements.
        \item Unique occupant configurations (e.g., no steering wheel, side-facing seats, or custom occupant restraint systems) could trigger additional reviews or standards, potentially adding multiple years to the timeline.
        \item States/provinces may also impose further hurdles (licensing, labeling, insurance requirements) that can delay widespread deployment.
    \end{itemize}
\end{itemize}

Particularly \emph{unconventional} AV designs (e.g., no pedals or steering wheel, non-traditional occupant positions, or brand-new “skateboard” vehicle platforms) face extra scrutiny. The unfamiliar geometry and crashworthiness implications often require extensive simulation, physical crash tests, and iterative design changes to gain approval. Consequently, \(\mathbf{T_{\mathrm{prod+reg}}}\) can be \emph{significantly} longer—on the order of 5--7 years or more for radical designs—compared to 3--5 years for relatively incremental updates (such as integrating Level 3 or Level 4 features into an existing passenger car platform).

In subsequent analyses, we incorporate \(\mathbf{T_{\mathrm{prod+reg}}}\) into the \emph{overall} autonomous-vehicle timeline. Even if an AV meets its safety (reliability) targets and has sufficient HPC resources, mass deployment may still be delayed by product-development cycles and regulatory reviews.

\subsection{Computational Feasibility and HPC Timelines (\texorpdfstring{$T_{\mathrm{comp}}$}{Tcomp})}

Recall from Section~\ref{sec:stateSpace} that we may need up to $10^{16}$ ops/s for real-time multi-agent planning. Current on-vehicle compute is around $10^{12}$--$10^{13}$ ops/s. Even with doubling every ~2--3 years, bridging several orders of magnitude may take 10+ years. Let $T_{\mathrm{comp}}$ be that horizon for adequate HPC (on-vehicle or distributed). Section~\ref{sec:techReadiness} further details how we model HPC growth rates, sensor improvements, and software maturity.

\subsection{Reliability Growth Timelines (\texorpdfstring{$T_{\mathrm{rel}}$}{Trel})}

Recall from Section \ref{sec:ReliabilityGrowth} that the reliability growth of an AV system can be modeled by the \emph{Crow-AMSAA} reliability growth model and the \emph{Poisson-based} safety validation model. Both models yield a required mileage \(R(t)\) in order to demonstrate that an AV’s failure rate has reached an acceptable threshold for deployment. Application of these models yields \(\mathbf{T_{\mathrm{rel}}}\) for a real-world calendar schedule. 

\subsection{Combining Timelines}

From the reliability side (Section~\ref{sec:ReliabilityGrowth}), we define $T_{\mathrm{rel}}$ (or the more refined $T_{\mathrm{rel}}$') as the time required to accumulate sufficient test miles for demonstrating an acceptably low failure rate. However, \textbf{meeting reliability targets alone is not enough}; we must also wait for HPC feasibility (\(T_{\mathrm{comp}}\)) and then add the product/regulatory cycle (\(T_{\mathrm{prod+reg}}\)):

\begin{equation}
\label{eq:combined_timeline}
T_{\mathrm{total}} \;=\;
\max\bigl(T_{\mathrm{comp}},\, T_{\mathrm{rel}}\bigr)
\;+\;
T_{\mathrm{prod+reg}}.
\end{equation}

\textbf{Note on HPC vs.\ Reliability Bottleneck:}  
If $T_{\mathrm{rel}}$' (our refined version of $T_{\mathrm{rel}}$) is \emph{shorter} than $T_{\mathrm{comp}}$, then HPC feasibility is the slower element, effectively gating the entire program. Conversely, if HPC matures quickly but reliability demonstration remains long, $T_{\mathrm{rel}}$ (or $T_{\mathrm{rel}}$') will be the bottleneck. Regardless, once the slower of these two completes, we still require $T_{\mathrm{prod+reg}}$ to finish product engineering and regulatory signoff. Hence, the final timeline for broad deployment in a given category is:

\begin{equation}
\label{eq:combined_timeline2}
T_{\mathrm{total}} \;=\;
\max\bigl(T_{\mathrm{comp}},\, T_{\mathrm{rel}}\bigr)
\;+\;
T_{\mathrm{prod+reg}}.
\end{equation}

In practical terms, even if $T_{\mathrm{rel}}$ is relatively short (because the AV accumulates billions of test miles per year), the vehicle may remain non-viable for widespread sale if $T_{\mathrm{comp}}$ (on-vehicle HPC) lags behind or if $T_{\mathrm{prod+reg}}$ stretches due to regulatory or engineering complexity. Similarly, if HPC rapidly advances but reliability demonstration remains slow, $T_{\mathrm{rel}}$ will be the gating factor.

In the following sections, we apply this combined timeline formula to illustrate how multi-decade horizons can emerge under realistic assumptions for each domain (consumer automotive, robo-taxis, highway trucking, etc.).

\subsection{Hybrid Serial--Parallel Timeline Model}
\label{sec:hybridSerialParallel}

In many real-world AV development programs, \emph{some} fraction of the reliability growth (i.e., finding and fixing issues) can proceed in parallel with evolving HPC hardware, while \emph{final} reliability improvements (especially on complex corner cases) may only begin once high-performance computing has reached a certain threshold. Also, once the system is “frozen,” a separate \emph{final QA} or \emph{Poisson-based zero-failures test} may be performed. 

To capture this reality, we can break the total timeline into partial overlaps plus final phases. Concretely, define two Crow-AMSAA times:
\begin{itemize}
    \item $T_{\mathrm{rel\mhyphen crow,\:partial}}$: The \textbf{initial, partial reliability-growth phase} that can happen even without full HPC maturity. We envision that up to some fraction of the environment (e.g., simpler roads, fewer edge cases) can be tested in parallel while hardware is still ramping up.
    \item $T_{\mathrm{rel\mhyphen crow,\:final}}$: The \textbf{final reliability-growth phase} that \emph{must} wait for HPC above a certain threshold in order to tackle the hardest scenarios (dense urban merges, very high-speed highways, extreme weather, etc.). 
\end{itemize}

Once all fixes and final system configurations are complete, a separate \emph{Poisson-based zero-failures test} (\S\ref{sec:QuantSafetyVal}) may be conducted to achieve confidence $C$ that the final system meets the target failure rate. Finally, product engineering and regulatory approval times $T_{\mathrm{prod+reg}}$ are appended.

\paragraph{Mathematical Expression.}
We define the total timeline $T_{\mathrm{total}}$ as:
\begin{equation}
\label{eq:hybridParallelSerial}
T_{\mathrm{total}}
\;=\;
\max\bigl(T_{\mathrm{rel\mhyphen crow,\:partial}},\; T_{\mathrm{comp}}\bigr)
\;\;+\;\;
T_{\mathrm{rel\mhyphen crow,\:final}}
\;\;+\;\;
T_{\mathrm{rel\mhyphen poisson}}
\;\;+\;\;
T_{\mathrm{prod+reg}},
\end{equation}
where:
\begin{itemize}
    \item $T_{\mathrm{rel\mhyphen crow,\:partial}}$ = time spent on reliability-growth testing \emph{in parallel} with HPC ramp-up, addressing simpler or mid-level scenarios.
    \item $T_{\mathrm{comp}}$ = HPC feasibility horizon (\S\ref{sec:techReadiness}), i.e.\ the time until HPC is mature enough to handle the \emph{most demanding} scenarios.
    \item $T_{\mathrm{rel\mhyphen crow,\:final}}$ = additional reliability-growth effort once HPC is available to test the advanced (most complex) edge cases. 
    \item $T_{\mathrm{rel\mhyphen poisson}}$ = final \emph{Poisson-based zero-failures demonstration} (\S\ref{sec:QuantSafetyVal}). One might consider this a short, dedicated “final QA” or “freeze” test.
    \item $T_{\mathrm{prod+reg}}$ = product engineering and regulatory approval (\S\ref{sec:prodReg}).
\end{itemize}
In other words, we allow partial reliability improvement to occur in parallel with HPC progress, but we do \emph{not} proceed to \emph{final} reliability closure or advanced scenario coverage until HPC passes its gating milestone. 

\paragraph{Narrative Explanation.}  
Some AV developers can indeed start discovering and fixing “ordinary” bugs long before the ultimate HPC hardware is ready; this is reflected by $T_{\mathrm{rel\mhyphen crow,\:partial}}$ happening in parallel with $T_{\mathrm{comp}}$. However, once HPC crosses a necessary performance threshold at time 
\(
  \max\bigl(T_{\mathrm{rel\mhyphen crow,\:partial}}, \, T_{\mathrm{comp}}\bigr),
\)
the team shifts focus to the most computationally demanding use-cases, requiring an \emph{additional} time $T_{\mathrm{rel\mhyphen crow,\:final}}$ to weed out advanced edge-case failures. Finally, after the entire system is stable and presumably “frozen,” a \emph{separate} Poisson-based zero-failures test ($T_{\mathrm{rel\mhyphen poisson}}$) is run to confirm a final $\lambda_{\mathrm{target}}$ with high confidence $C$. Product engineering cycles and regulatory reviews $T_{\mathrm{prod+reg}}$ are then appended.

\paragraph{Estimating Partial vs.\ Final Crow-AMSAA Durations.}
One possible approach:
\begin{enumerate}
    \item \textbf{Crow-AMSAA total:} First compute $T_{\mathrm{rel\mhyphen crow,\:total}}$ from your usual power-law approach, i.e.\ 
    \(
        T_{\mathrm{rel\mhyphen crow,\:total}} = \frac{R_{\mathrm{crow}}}{M},
    \)
    with $R_{\mathrm{crow}}$ from Eq.~(\ref{eq:R_CrowAMSAA}). 
    \item \textbf{Partial fraction:} Decide (based on engineering judgment) that a fraction $f$ (like 70\% or 80\%) of that reliability-improvement mileage can be done without final HPC. The partial time is then 
    \[
    T_{\mathrm{rel\mhyphen crow,\:partial}} = f \times T_{\mathrm{rel\mhyphen crow,\:total}}.
    \]
    \item \textbf{Final fraction:} The remaining $(1 - f)$ fraction happens after HPC is mature:
    \[
    T_{\mathrm{rel\mhyphen crow,\:final}} = (1 - f)\;\times\;T_{\mathrm{rel\mhyphen crow,\:total}}.
    \]
    \item \textbf{Final Poisson QA:} If separate, pick $T_{\mathrm{rel\mhyphen poisson}}$ from your zero-failures formula or the typical mileage you plan to run in final QA. 
\end{enumerate}
In practice, $f$ might be 0.7 or 0.8 if you believe most non-trivial reliability fixes can be found with moderate HPC, reserving just 20--30\% for the hardest corner cases that truly require HPC extremes.

\paragraph{Example.}  
If $T_{\mathrm{comp}}=15$~yrs, $T_{\mathrm{rel\mhyphen crow,\:total}}=10$~yrs, $f=0.7$, and $T_{\mathrm{rel\mhyphen poisson}}=1$~yr, then
\[
T_{\mathrm{rel\mhyphen crow,\:partial}} = 0.7 \times 10 = 7\;\text{yrs},
\quad
T_{\mathrm{rel\mhyphen crow,\:final}}= 3\;\text{yrs}.
\]
Hence
\[
\max\bigl(T_{\mathrm{rel\mhyphen crow,\:partial}},\,T_{\mathrm{comp}}\bigr) 
= \max(7,\,15)=15,
\quad
\]
\text{then add } \[
T_{\mathrm{rel\mhyphen crow,\:final}}=3 \text{ and } T_{\mathrm{rel\mhyphen poisson}}=1
\text{ plus } T_{\mathrm{prod+reg}}.
\]

So overall 
\[
T_{\mathrm{total}} = 15 + 3 + 1 + T_{\mathrm{prod+reg}} = 19 + T_{\mathrm{prod+reg}}.
\]
If $T_{\mathrm{prod+reg}}=4$, total is 23~yrs. This approach yields an intuitive “hybrid” result: partial reliability tasks occur \emph{in parallel} with HPC ramp-up, but advanced tasks must \emph{wait} for HPC to cross its gating milestone. 

\paragraph{Complete Timeline Formula Including \boldmath$f$.}
To summarize the above approach more explicitly, let
\[
T_{\mathrm{rel\mhyphen crow}}
\;=\;
\frac{\,R(t)\,\times\,\gamma\,\times\,\delta\,}%
     {\,\text{M}\,\\},
\]
and choose a fraction \(0 \le f \le 1\) such that \(f\,T_{\mathrm{rel\mhyphen crow}}\) may run \emph{in parallel} with HPC maturation, while the remaining \((1 - f)\,T_{\mathrm{rel\mhyphen crow}}\) must occur after HPC is fully ready.  Then the total timeline becomes:
\begin{equation}
\label{eq:total_with_f}
\boxed{%
T_{\mathrm{total}}
\;=\;
\max\!\Bigl(\,f\,T_{\mathrm{rel\mhyphen crow}},\,T_{\mathrm{comp}}\Bigr)
\;+\;
\bigl(1-f\bigr)\,T_{\mathrm{rel\mhyphen crow}}
\;+\;
T_{\mathrm{rel\mhyphen poisson}}
\;+\;
T_{\mathrm{prod+reg}}.}
\end{equation}
In this manner, the user can tune \(f\) to reflect how much of the Crow-AMSAA mileage or reliability growth can realistically be accomplished without final HPC resources. When \(f=0\), \emph{all} reliability growth is deferred until after HPC readiness; when \(f=1\), HPC is effectively \emph{not} a gating item, and the reliability curve proceeds entirely in parallel. For most realistic AV scenarios, \(f\) is between 0.5 and 0.8, indicating partial concurrency followed by additional HPC-driven final testing.

\section{Stages of Deployment}
\label{sec:StagesDeployment}

In this section, we present a structured \textbf{3-stage} view of autonomous vehicle deployment that underpins all subsequent timeline projections. We use these stages to categorize the maturity and scale of AV operations:

\begin{itemize}
    \item \textbf{Stage 1 (Pilot \& Initial Limited Deployment):}\\ 
    AV operation is confined to narrow or geofenced domains, often with a safety driver or remote human fallback. Vehicles at this stage are typically custom-outfitted or retrofitted prototypes. Though publicly visible, these pilots run in limited service and often collect data to validate core functionality. The operational environment may be simplified (e.g., low-speed loops, fair weather only), and reliability requirements are less strict than full commercial service.
    
    \item \textbf{Stage 2 (Revenue Service Deployment):}\\
    At this stage, AVs achieve reasonably stable operation in an expanded operational domain (ODD), and minimal or no remote interventions are typically needed under normal conditions. Operators may charge the public for rides or for delivery services, marking a shift to commercial viability. Vehicles are still often outfitted or retrofitted in low volumes; however, reliability targets tighten (relative to Stage 1), and regulatory oversight increases (city-by-city or corridor-by-corridor approvals).
    
    \item \textbf{Stage 3 (Broad Commercial Adoption):}\\
    This final stage covers near-universal deployment within the intended vehicle domain, with mass production and mainstream regulatory certification comparable to conventional vehicles. The AV system meets very high reliability thresholds, has scaled manufacturing (often by major OEMs or large-scale partners), and the service is cost-competitive with human-driven alternatives. No fallback driver or routine remote supervision is required, and the vehicle is broadly available to consumers or commercial fleets.
\end{itemize}

Sections~5.1--5.4 delve further into:
\begin{itemize}
    \item The specific \textbf{failure-rate thresholds} demarcating each stage,
    \item How \textbf{production and regulatory timelines} evolve from quick pilot waivers to full-scale factory certification,
    \item Why each stage typically has distinct HPC needs, reliability targets, and operational complexities.
\end{itemize}

\subsection{Failure-Rate Thresholds}

Recall that we can define numeric thresholds on a \emph{per hour} basis:

\begin{itemize}
    \item \textbf{Stage 1 (Pilot/Limited Deployment):} $P(\mathrm{failure}) < 10^{-7}$ per operating hour. Typically has safety operators onboard, restricted ODD, and possibly remote supervision. 
    \item \textbf{Stage 2 (Revenue Service Deployment):} $P(\mathrm{failure}) < 10^{-8}$ per operating hour. Operation is relatively stable, may charge the public for rides, and requires \emph{minimal to no remote interventions} under most normal conditions. 
    \item \textbf{Stage 3 (Broad Commercial Adoption):} $P(\mathrm{failure}) < 10^{-9}$ per operating hour. Near-universal coverage in that vehicle category, cost-competitive with humans, with little or no fallback driver or remote oversight.
\end{itemize}

\paragraph{Why These Thresholds?} 
They reflect orders-of-magnitude improvements in safety as we move from pilot projects (often considered “acceptable risk” if an operator is present) to large-scale commercial deployment needing extremely low probability of critical failures.

\subsection{Stage 1: Pilot and Initial Limited Deployment}

Stage 1 typically involves \textbf{custom-outfitted} or \textbf{retrofit} vehicles, limited to small or geofenced ODDs, often with a human safety driver or remote fallback. The pilot may be partially open to the public, or for testing/demonstration only. Because these are small-scale or even one-off prototypes:

\begin{itemize}
    \item \textbf{Production Time:} Generally 6--12 months to outfit an existing vehicle platform with sensors, compute hardware, etc.
    \item \textbf{Regulatory Time:} 1--6 months to obtain waivers or operate under limited legal boundaries. 
    \item \textbf{Exception (Bespoke Shuttles):} For truly bespoke (aka purpose-built) low-volume shuttles, building even a handful of prototypes can take 2--3 years, plus another 6--12 months for waivers. 
\end{itemize}

Hence, \emph{from a timeline perspective}, Stage 1 is already happening for most categories, but the time we compute for Stage 1 in Section~7 references \textbf{when these pilots become robust enough to truly validate} that $10^{-7}$/hr threshold (leading to potential moves toward Stage 2).

\subsection{Stage 2: Revenue Service Deployment}

In Stage 2, the AV services are \textbf{stable enough to charge revenue}, with minimal or no remote oversight in most normal operations. This implies an \emph{improved} reliability requirement ($10^{-8}$/hr) and typically an \emph{expanded} ODD. However, many vehicles in Stage 2 are still \textbf{custom-outfitted or retrofitted} (particularly if overall volume is still small):

\begin{itemize}
    \item \textbf{Production Time:} Similarly 6--12 months for outfitting existing vehicles. 
    \item \textbf{Regulatory Time:} 1--6 months for waivers or city-by-city legal acceptance. 
    \item \textbf{Exception (Bespoke Shuttles):} Still might require 2--3 years to build small production runs and 6--12 months for limited approvals, especially if occupant seating is unconventional.
\end{itemize}

Thus, even though Stage 2 is \emph{commercially} oriented (the public pays for a ride, or the vehicle runs goods routes reliably), the regulatory environment is still patchwork. Large-scale from-factory production is not yet the norm.

\subsection{Stage 3: Broad Commercial Adoption}

In Stage 3, $P(\mathrm{failure}) < 10^{-9}$/hr represents \textbf{broad, near-universal coverage} within the vehicle’s intended domain, and cost-competitive with human operation. Here, we assume:

\begin{itemize}
    \item \textbf{Production/Reg Time:} The longer cycles discussed in Section~\ref{sec:prodReg} (3--5 years, or up to 5--7 for bespoke purpose-built designs) now apply. That’s because the vehicle is typically \textbf{outfitted at large scale from a factory} (OEM or major supplier + software integrator), requiring mainstream crashworthiness certification, occupant protection designs, and regulatory sign-off. 
    \item \textbf{Highly integrated HPC and sensor stacks} rather than bolt-on retrofits.
    \item \textbf{Mature reliability demonstration,} ensuring near-zero reliance on fallback drivers or frequent remote interventions.
\end{itemize}

Accordingly, Stage 3 demands the full synergy of HPC readiness, advanced reliability growth, and extensive regulatory acceptance. This is why AV timeline calculations will tend to push out.

\section{Current Deployment Landscape}
\label{sec:CurrentPilots}

While the timelines for fully universal L5 remain unknown, we already see numerous \emph{ongoing} pilot and limited deployments that fit \textbf{Stage 1} criteria in many respects, and in some cases early revenue service that begin to approach \textbf{Stage 2}:

\begin{itemize}
    \item \textbf{Consumer Automotive (L2/3)}: Public road L2/L3 operations. Minimal L4 tests on public roads (e.g.\ Tesla FSD Beta, GM Super Cruise).
    \item \textbf{Robo-Taxis}: Waymo, Cruise geofenced areas; expansions often halted or scaled back due to incidents. 

    \begin{itemize}
    \item Note: \emph{Waymo} has been charging fares to the public in certain regions since ~2022, and by late 2024 offers more robust rides without human drivers onboard in some areas. These operations represent an \textbf{early revenue-service} deployment, but still involve significant oversight and remote tele-operations. Hence they are \emph{not} fully Stage 2 as defined in this paper, because they lack the minimal-to-no-remote-intervention standard and remain subject to narrow ODD restrictions and significant city-by-city overhead/waivers.
    \end{itemize}

    \item \textbf{Geo-Fenced Transit Vans/Buses}: Low to moderate speed fixed routes on campuses, corporate sites, airports, and urban transit locations.
    \item \textbf{Highway Trucking}: Platooning and on road tests; safety drivers remain onboard.
    \item \textbf{Delivery Vans}: Small suburban pilots, teleoperation fallback for tricky situations.
    \item \textbf{Bespoke Shuttles}: (aka purpose-built shuttles) Low-speed demos; hampered by custom vehicle costs, slow speeds, accidents, and waivers needed for public road operations.
    \item \textbf{Military/Defense}: Off-road prototypes, base-limited. Different cost/benefit calculus.
    \item \textbf{Industrial/Mining}: Automated haul trucks, dozers in fenced sites. Partial autonomy accelerating.
\end{itemize}

Thus, although we do see limited \emph{commercial} activity (like Waymo’s fare-charging robo-taxis), it still relies on \emph{significant operational constraints and remote monitoring}, which keeps it closer to Stage 1 in this paper’s framework. Table~\ref{tab:pilotstable} summarizes current deployments:

\FloatBarrier  

\begin{table}[h!]
\centering
\resizebox{\textwidth}{!}{%
\begin{tabular}{l|p{6cm}|p{6cm}|l}
\hline
\textbf{Category} & \textbf{Recent Examples} & \textbf{Status} \\
\hline
Consumer Automotive & L2/L3, limited L4 & Mostly partial automation \\
Robo-Taxis & Waymo and Cruise (Phoenix, SF) & Early service, heavy oversight \\
Geo-Fenced Vans/Buses & Fixed routes/loops & Operator onboard, fixed ODDs  \\
Highway Trucking & Platooning and corridor tests & High liability, slow progress  \\
Delivery Vans & Suburban pilots, teleop & Limited scale  \\
Bespoke Shuttles & Low-speed demos & Poor performance, costs, waivers  \\
Military/Defense & Off-road prototypes & Specialized R\&D pilots  \\
Industrial/Mining & Dozers, haul trucks & Partial autonomy in fenced sites  \\
\hline
\end{tabular}
}
\caption{Current Deployment Status (By Category)}
\label{tab:pilotstable}
\end{table}

\FloatBarrier  

\section{Category-Specific Timeline Calculations}
\label{sec:CategorySpecificTimelines}

This section applies the models outlined in this paper to different categories of autonomy across each \textbf{stage}. In the following subsections, we describe the rationale behind the selected parameter values used in the calculations, along with critical commentary. Note that we use values outlined in this paper (based on heuristics and guidelines from existing literature). These are not absolute, but rather parameters within a realistic realm, intended to guide us toward \textit{rough estimates} for relative AV timelines.

Also note that our focus is on calculating the AV timeline for \textbf{broad commercialization (Stage~3)}. However, we will on occasion speak to the \textbf{Stage~2} timelines that precede Stage~3. Appendix~A provides a table of the calculations made in this section.

\subsection{Target Compute for Naive MAPF}
\label{sec:7.1}

For the number of objects used in multi-agent path finding, \(n\), we have already asserted a baseline value of 60 as a conservative estimate for the most complex ODDs.

\begin{itemize}
    \item The most complex ODD being consumer automotive, we select \(n = 60\).
    \item For robo-taxis (a subset of consumer automotive in urban environments), we select \(n = 55\).
    \item For geofenced vans/buses, delivery vans, and bespoke shuttles---all operating at generally slower speeds and in more restricted and controlled environments---we select \(n = 35\).
    \item For military/defense ODDs, we also assume a definable operation (albeit sometimes offroad as well as on-road), with \(n = 35\).
    \item Finally, for highway trucking and industrial/mining applications (which are either pure highway operations or controlled work zones), we select \(n = 25\).
\end{itemize}

These \(n\) values allow us to calculate the target compute for naive MAPF across applications. With a real-time cycle of 100\,ms, we can thus compute the target compute demand for naive MAPF, \(C_{d}\).

\subsection{Real World Compute Demand Reductions}
\label{sec:7.2}

For practical consideration of target compute needs, in order to compute the \emph{effective} target compute demanded, \(C'_{d}\), we consider real-world reduction factors. For simplicity here, we list these as the set:
\[
\left\{
\begin{array}{l}
p_{1} = \text{limit for active objects reduction},\\
p_{2} = \text{temporal slicing reduction},\\
p_{3} = \text{distant agent reduction},\\
p_{4} = \text{local vs.\ global planning reduction},\\
p_{5} = \text{ODD restriction reduction}
\end{array}
\right\}.
\]

The factors considered per category (per stage) and compute reduction factor are shown in Appendix~A. Commentary on the computed reduction factor is presented here:

\begin{itemize}
    \item \textbf{Consumer auto} has the largest compute demand among the categories, followed by \textbf{robo-taxis}.
    \item \textbf{Geo-fenced vans/buses, delivery vans, bespoke shuttles, and military/defense} applications are assumed to have approximately the same reduction levels.
    \item Given the faster speeds of \textbf{highway trucking}, the reduction levels are assumed to be less than consumer auto but more than the slower-speed geo-fenced applications.
    \item \textbf{Industrial/mining} applications are assumed to leverage the lowest compute demand among the categories, given relatively sparse and controlled ODDs plus generally slower speeds.
\end{itemize}

The resulting reduced compute demand, \(C'_{d}\), can then be calculated using these reduction heuristics. With a current compute capability of \(C_{c}\) (as assumed in the paper) and Moore’s Law--driven \(T_{d} = 2.5\) (also described), we can compute \(T_{\mathrm{comp}}\) for the various categories. If the compute demanded is less than what is currently available, we assume 0~years for \(T_{\mathrm{comp}}\), with the understanding that other factors may bound us instead.

\subsection{Target Compute Demand Timelines}
\label{sec:7.3}

For all categories \emph{except} consumer auto and robo-taxis, we calculate \(T_{\mathrm{comp}}\) as 0, meaning that the compute power available today in embedded form is presumably sufficient to meet the needs of those applications. Thus, due to reduced speeds, ODD limitations, or relative ODD simplicity, the compute available for geo-fenced vans/buses, highway trucking, delivery vans, bespoke shuttles, military/defense, and industrial/mining is presumed sufficient. This aligns with real-world observations for such applications.

However, for consumer auto and robo-taxis, to meet the demands required for \textbf{Level~5 autonomy} across a wide range of conditions and edge-case processing for broad commercialization, the required compute is presumed more significant. With numbers such as \(\sim20\)~years out for robo-taxis and \(\sim35\)~years out for consumer auto, we are left to ponder the reality of these results and our parameter selections. Yet these numbers are not surprising from another perspective: the complexity of these environments has often been underestimated. We are only at the cusp of discovering edge cases as pilot deployments (and even early revenue-generating operations) proceed. How much further do we have to go? If we look to the complexity of objects that drive compute demand, and similarly assume parallel timeframes for software technology readiness to exploit this compute, then such timeframes may not be far off the mark.

Reduced ODD revenue-generation applications may, however, drive the compute demand down. Nevertheless, for driverless Stage~2 revenue service with minimal or no remote intervention, we might still project \(\sim15\)~years for robo-taxis and \(\sim25\)~years for consumer auto to achieve sufficient compute for a full-bloom Stage~2 deployment.

\subsection{Crow-AMSAA Constants}
\label{sec:7.4}

For calculation of Crow-AMSAA--based timelines, we chose to leverage the same values for \(\alpha\) and \(\beta\) across all categories. Since \(\alpha\) is an empirically fitted constant (and can vary across timelines), we choose \(\alpha = 0.0001\). The rationale is that the autonomous vehicle industry is now in full swing; furthermore, the programs seriously pursuing autonomy will drive the timeline to broad commercialization. Since we cannot easily distinguish \(\alpha\) across categories (it is organization-dependent), we use a factor from the paper and apply it consistently across categories for Stage~3 L5 broad commercialization.

Similar arguments hold for the selection of the reliability growth exponent, \(\beta\). Because \(\beta\) can strongly affect timelines, we chose \(\beta = 0.4\) (driven by the paper’s exposition on the underlying model) and applied it consistently. The \(\lambda_{\mathrm{target}}\) factor we used is one fatality incident per \(10^{8}\) miles, roughly equivalent to human-level (and in fact somewhat smaller than the \(1.42\times10^{-8}\) “human factor”).

Finally, recall that the factor \(f\) drives which percentage of testing occurs even while HPC target demand is being met (\(T_{\mathrm{crow,partial}}\)) and which percentage of testing occurs once HPC target compute is achieved (\(T_{\mathrm{crow,final}}\)). For fairness across categories, we ascribe the same constant from our paper’s narrative: \(f = 0.7\). Thus, we assume 70\% of testing can occur while HPC target demand is being met, and then once met, an additional 30\% of test time is required to complete the bounds of \(T_{\mathrm{crow}}\).

\subsection{Severity Level}
\label{sec:7.5}

The severity level \(s\) is proportional to the number of fatalities likely in an incident where fatalities occur.

\begin{itemize}
    \item For \textbf{consumer auto, geo-fenced vans/buses, and delivery vans}, we choose a baseline severity of \(s = 1\).
    \item For \textbf{highway trucking}, we expect a larger severity than that of lower-speed, smaller-mass vehicles. A single tractor-trailer at highway speed can cause a mass-casualty incident. Even if only one or two vehicles are impacted, the severity is high, so we estimate \(s = 5\).
    \item For \textbf{military/defense} vehicles, multiple casualties could exist, but these are inherently risky domains, so we level the severity to \(s = 1\).
    \item The same applies to \textbf{industrial/mining} applications, where there may be no humans in proximity. We choose a baseline \(s = 1\).
    \item \textbf{Robo-taxis}, by their nature (carrying one or more passengers in busy urban settings with pedestrians), are given \(s = 2\).
    \item \textbf{Bespoke shuttles}, because of their relatively untested platforms and crashworthiness, plus their function akin to a robo-taxi, are also set to \(s = 2\) for our purposes.
\end{itemize}

\subsection{Test Miles}
\label{sec:7.6}

The paper’s \(M\) indicates the annual testing capacity (in miles) for an AV program. These miles may be real or simulated. Inevitably, some real miles are needed, but to meet scenario-variance demands, some number of simulated miles are typically required too. Since we cannot surmise exactly what any given company will do, or how it might vary by category, we assume a \emph{similar} value across categories. Moreover, given the push for autonomy, we assume this number is necessarily large for programs aiming at broad L5 commercialization. Hence, we set
\[
M = 10^{9} \quad (\text{a billion miles per year}),
\]
through some combination of real-world and simulated testing.

Even with \(10^{9}\) annual test miles, the timelines (as we shall see) are significant. If we were even one order of magnitude lower, a resultant 10\(\times\) increase in time would push already-long timeframes further. Hence, to be successful, this figure also serves as a guidepost for the magnitude of test mileage that might be required for broad L5 commercialization (Stage~3). For Stage~2 revenue service and more restricted ODDs, this number can be significantly smaller, enabling earlier Stage~2 deployments.

\subsection{ODD Reduction and Complexity}
\label{sec:7.7}

Regarding the \(\delta\) variable that describes ODD reduction factors, the scope of an ODD may be assumed cut in half for Stage~2 revenue service (or even less). But for Stage~3 L5 broad commercialization, we assume \(\delta = 1.0\), meaning \emph{no ODD reduction} in achieving full broad L5 coverage.

Meanwhile, \(\gamma\) is the ODD \textit{complexity} factor, varying by category:

\begin{itemize}
    \item For \textbf{consumer auto} (the most complex ODD), we start with \(\gamma = 1\).
    \item \textbf{Robo-taxis} are close behind because of urban environments, but often not the entire scope of consumer auto, so \(\gamma = 0.9\).
    \item \textbf{Geo-fenced vans/buses, delivery vans, bespoke shuttles} all share similar ODD constraints, so for this paper we use \(\gamma = 0.5\), about half as complex as consumer auto. In reality, this number might evolve from as low as 0.1 up to this presumed peak for L5 broad commercialization.
    \item \textbf{Highway trucking}, while higher speed, can be simpler from a geometry/environment perspective. Thus, we choose \(\gamma = 0.4\).
    \item \textbf{Military/defense} applications, with fixed missions and accepted risk domains, are assumed to have \(\gamma = 0.3\).
    \item Finally, \textbf{industrial/mining} applications presumably have the simplest ODD (little human interaction, well-bounded work zones), resulting in \(\gamma = 0.2\).
\end{itemize}

\subsection{Crow-AMSAA Timelines}
\label{sec:7.8}

Given these factors, we can calculate the projected timeline from the Crow-AMSAA model across categories:

\begin{itemize}
    \item A \textit{low severity} and \textit{low ODD complexity} drive smaller required test-mile counts which, combined with the standard assumed possible across categories, yield the smallest \(T_{\mathrm{crow}}\) for \textbf{industrial/mining} and \textbf{military/defense applications} (on the order of 2--3~years). In reality, lesser real-world testing capacity or more complex ODDs might lengthen these times, but for this paper’s relative comparison we see \(\sim2\text{--}3\) years.
    \item \textbf{Geo-fenced vans/buses and delivery vans} share a similar timeline, with moderate \(\gamma\), resulting in \(\sim5\) years for \(\,T_{\mathrm{crow}}\).
    \item \textbf{Bespoke shuttles} and \textbf{robo-taxis} have a higher severity. Their total miles demanded under this model thus increase significantly over geo-fenced applications. Bespoke shuttles’ ODD complexity is almost half that of robo-taxis, yet the final horizon for both categories to achieve L5 broad commercialization is projected much further. The calculation here is \(\sim30\) years for bespoke shuttles and \(\sim50\) years for robo-taxis. These are large numbers, but we will see how they play into the final calculations.
    \item For \textbf{consumer auto}, because we set a “blended” severity baseline, we get a more moderate (though still nontrivial) projection of \(\sim10\) years for broad L5 commercialization.
    \item For \textbf{highway trucking}, the timeline modeled is extremely long (\(225{+}\) years), driven by the severity assumption (\(s = 5\)). One could argue this suggests an application that is not truly viable at L5. Or it may imply we need disruptive breakthroughs (e.g.\ dedicated freight lanes) to render this timeline more practical.
\end{itemize}

We acknowledge that these results might seem extreme. Quantitative modeling can yield such large numbers for high-severity applications, indicating that either (1) the model is overly conservative or (2) these applications face truly daunting reliability requirements. Future study can refine these assumptions, but it is worth noting that high-severity autonomous vehicles may remain quite distant or require fundamentally different risk mitigation strategies.

\subsection{Poisson Timelines}
\label{sec:7.9}

For our Poisson-based timeline calculations (Section~3.2 of the paper), we choose a confidence \(C = 0.95\) and a safety factor \(\mathrm{SF} = 2.0\) across the board. We also choose \(\lambda_{\mathrm{target}} = 7.1 \times 10^{-9}\), which is 50\% better than the human baseline of \(1.42 \times 10^{-8}\). Using the same test miles, \(\gamma\), and \(\delta\) for \(T_{\mathrm{crow}}\) calculations, we arrive at \(T_{\mathrm{poisson}}\) figures that approximate the time needed for final QA with a stable system, \emph{after} the Crow-AMSAA testing. Choosing these common constants yields \(\sim1\) solid year of uninterrupted testing to meet that target failure rate with confidence \(C\) and safety factor \(\mathrm{SF}\).

Applying the same ODD reduction (\(\delta\)) and complexity (\(\gamma\)) numbers here as in the \(T_{\mathrm{crow}}\) calculations \emph{further reduces} the Poisson-based QA test time required to achieve the necessary confidence at the permitted failure rates. Concretely, this yields:

\begin{itemize}
    \item \textbf{2--3 months} for \emph{industrial/mining} and \emph{military/defense} domains,
    \item \textbf{4--5 months} for \emph{geo-fenced vans/buses}, \emph{delivery vans}, \emph{bespoke shuttles}, and \emph{highway trucking},
    \item \textbf{9--10 months} for \emph{robo-taxis} and \emph{consumer automotive}.
\end{itemize}

\subsection{Production and Regulatory Timelines}
\label{sec:7.10}

Although it might be argued these timelines can happen in parallel with some of the other intervals, for \emph{massive scale} and \textbf{L5 broad commercialization}, one usually needs a stable underlying AV design. 

\begin{itemize}
    \item For \textbf{consumer auto}, typical 3--5~year production cycles apply once the design is stable for mass production. Adding regulatory approvals (which large OEMs are well-equipped to handle), we assume \(\sim5\) years total.
    \item Since \textbf{robo-taxis} and \textbf{highway trucking} generally build on commercially available vehicles, we also assume \(\sim5\) years for production plus regulatory.
    \item \textbf{Bespoke (purpose-built) shuttles}, on the other hand, require more extensive crashworthiness testing and novel platform approvals. We assume \(\sim7\) years. In reality, it could be longer, given the novelty of some platforms.
    \item \textbf{Geo-fenced vans/buses, delivery vans, military/defense vehicles, and industrial/mining} vehicles are all assumed to have shorter product/approval cycles (2.5~years) because they are integrated with or build on existing platforms and generally operate under more constrained ODDs.
\end{itemize}

\subsection{Calculating Total AV Timelines}
\label{sec:7.11}

From the model in Section~4.5 of this paper, we have:
\[
T_{\mathrm{total}} 
= \max\bigl(T_{\mathrm{rel\mhyphen crow, partial}},\; T_{\mathrm{comp}}\bigr) 
\;+\; 
T_{\mathrm{rel\mhyphen crow, final}} 
\;+\; 
T_{\mathrm{rel\mhyphen poisson}} 
\;+\;
T_{\mathrm{prod+reg}}.
\]
Using the year~\textbf{2024} as our baseline and adding the total to 2024 gives us a \textit{rough} projected date for achieving L5 broad commercialization.

Pulling the calculations from above (and expanded in Appendix~A), we get total estimates for the time required across categories:

\begin{itemize}

    \item \textbf{Industrial/Mining:} 
    Among the earliest timelines are industrial/mining applications, with a total AV timeline of \(\sim5\) years for broad L5 commercialization (Stage~3). 
    While partial revenue service is already occurring, their full-on Stage~2 revenue service level is \(\sim2\) years away. 
    \begin{itemize}
        \item (Stage~2 = 2026, Stage~3 = 2029)
    \end{itemize}

    \item \textbf{Military/Defense:} 
    Close on the heels of industrial/mining are defense/military applications, with a total AV timeline of \(\sim6\) years to Stage~3 and \(\sim3\) years to Stage~2. 
    Although partial revenue service is under way, the path to full-scale operations still requires additional testing and reliability growth. 
    \begin{itemize}
        \item (Stage~2 = 2027, Stage~3 = 2030)
    \end{itemize}

    \item \textbf{Delivery Vans:} 
    With simplified ODDs, delivery vans have a near-horizon commercialization target of \(\sim8\) years for broad L5. 
    Stage~2 revenue service is projected at \(\sim4\) years, noting that partial revenue service and pilot programs have already begun. 
    \begin{itemize}
        \item (Stage~2 = 2028, Stage~3 = 2032)
    \end{itemize}

    \item \textbf{Geo-fenced Vans/Buses:}
    Geo-fenced passenger vans and buses are similarly under way, sharing the reliability and scale-growth path of delivery vans (due to similar ODDs). 
    Stage~3 commercialization is projected at \(\sim8\) years, with Stage~2 revenue service \(\sim4\) years out. 
    \begin{itemize}
        \item (Stage~2 = 2028, Stage~3 = 2032)
    \end{itemize}

    \item \textbf{Bespoke Shuttles:} 
    Bespoke (purpose-built) shuttles have a significantly longer path to both safe revenue service and broad commercialization compared to their geo-fenced counterparts. 
    Driven by the demands of the reliability growth model, higher severity levels, and additional production and regulatory time, this model yields \(\sim35\) years for Stage~3 and \(\sim20\) years for safe Stage~2 revenue service. 
    The long timeframe partly reflects the novelty of these platforms, which may need to be proven safe even for manual operations before fully autonomous service. 
    \begin{itemize}
        \item (Stage~2 = 2043, Stage~3 = 2060)
    \end{itemize}

    \item \textbf{Consumer Automotive:} 
    Consumer auto faces extensive edge cases, high ODD diversity, and major compute/technology readiness hurdles. 
    Combining production and regulatory cycles plus the needed time for further testing, we end up with \(\sim30\) years to Stage~2 revenue commercialization and \(\sim43\) years to Stage~3. 
    While this may seem surprising given the massive industry push, current “auto-drive” features are primarily advanced ADAS, and the long tail of ODD variety remains a major factor. 
    \begin{itemize}
        \item (Stage~2 = 2054, Stage~3 = 2067)
    \end{itemize}

    \item \textbf{Robo-Taxis:} 
    With even more demanding complexity (dense urban ODDs, higher severity), robo-taxis appear further out than consumer auto. 
    This is partly explained by a lengthier Crow-AMSAA reliability growth horizon. 
    Although robo-taxis occupy a subset of the consumer auto ODD, they may actually precede full consumer auto in Stage~2 deployments once certain urban domains are mastered. 
    \begin{itemize}
        \item (Stage~2 = 2051, Stage~3 = 2081)
    \end{itemize}

    \item \textbf{Highway Trucking:} 
    We do not know whether these models fully capture automated highway trucking or if they reveal the extreme realities of this domain. 
    The high severity, higher speeds, and heavy vehicle mass push the timeline off the charts—\(\sim114\) years for Stage~2 and \(\sim223\) years for Stage~3. 
    It may be that highway trucking will require a disruptive breakthrough (e.g.\ dedicated freight lanes) to be truly viable. Future work can further evaluate these factors. 
    \begin{itemize}
        \item (\emph{Off charts:} Stage~2 = 2138, Stage~3 = 2253)
    \end{itemize}

\end{itemize}

While pilots exist in all these categories (and some already have partial revenue service), that is still limited and restricted. The timelines for \textbf{full} Stage~2 revenue commercialization, with minimal or no intervention, stretch further. And Stage~3 broad L5 commercialization---the ultimate driverless capability meeting this paper’s safety and reliability marks---may be \textit{much} further out, despite marketing claims to the contrary.

\section{Results}
\label{sec:results}

In this section, we provide a concise summary of the outcomes derived in Section~7. Table~\ref{tab:summaryTable2} lists the estimated dates for both \textbf{Stage~2 Revenue Service} and \textbf{Stage~3 Broad Commercialization} across the various AV categories, ordered as introduced in Section~7.1.

\vspace{1em}
\begin{table}[h!]
\centering
\renewcommand{\arraystretch}{1.15}
\resizebox{\textwidth}{!}{
\begin{tabular}{lcc}
\hline
\textbf{Category} & \textbf{Revenue Service (Stage 2)} & \textbf{Broad Commercialization (Stage 3)}\\
\hline
\textbf{Industrial/Mining}     & 2026 & 2029 \\
\textbf{Military/Defense}      & 2027 & 2030 \\
\textbf{Delivery Vans}         & 2028 & 2032 \\
\textbf{Geo-fenced Vans/Buses} & 2028 & 2032 \\
\textbf{Bespoke Shuttles}      & 2043 & 2060 \\
\textbf{Consumer Automotive}    & 2054 & 2067 \\
\textbf{Robo-Taxis}            & 2051 & 2081 \\
\textbf{Highway Trucking} (\emph{Off Charts})      & \emph{2138} & \emph{2253} \\
\hline
\end{tabular}
}
\caption{Estimated Timeline for Stage~2 \& Stage~3 by Category}
\label{tab:summaryTable2}
\end{table}

\vspace{1em}

\noindent
\textbf{Interpretation of the Results.} \\
The data in Table~\ref{tab:summaryTable2} reflect each category’s estimated timeline to:
\begin{itemize}
    \item \emph{Stage~2 Revenue Service}, where AVs can operate with minimal (if any) remote supervision, and
    \item \emph{Stage~3 Broad Commercialization}, representing near-universal deployment within the intended domain, where the AV is fully driverless at human-competitive reliability and cost.
\end{itemize}

Several overarching points stand out:

\begin{enumerate}
\item \textbf{Categories with Limited ODD or Fewer Interactions Appear Soonest.}\\
Industrial/mining and military/defense operations already see partially automated vehicles in production or testing. Because these domains are inherently restricted or controlled, the complexity (and thus required reliability threshold) can be met sooner.

\item \textbf{Consumer-Focused Categories Require Longer Horizons.}\\
For example, consumer automotive and robo-taxis must handle extremely diverse road conditions and a vast number of edge cases. These factors inflate the required testing and computational capacity, pushing Stage~2 and Stage~3 timelines further.

\item \textbf{High Severity Leads to Even Longer Timelines.}\\
Highway trucking stands out as an exceptionally large figure, driven by the severity factor (due to mass and speeds), which the reliability models interpret as requiring enormous test mileage and a much longer horizon unless there is a structural or infrastructural disruption (e.g., dedicated autonomous freight lanes).

\item \textbf{Broad L5 Commercialization Takes Decades for Some.}\\
Bespoke shuttles, consumer cars, and robo-taxis all show multi-decade timelines. This reflects either high complexity (e.g., urban environments) or the novelty of the vehicle platforms (as in the case of bespoke shuttles) where both reliability-growth and production/regulatory cycles must be satisfied.

\item \textbf{Parallel Versus Serial Development.}\\
Although these numbers look large, much of the reliability and HPC ramp-up (see Section~4.5) occurs in parallel. Early pilots and restricted ODD deployments still evolve. The final steps—\emph{broad} Stage~3 rollouts—require near-zero interventions, high confidence, and stable hardware/software platforms.

\end{enumerate}

\noindent
Overall, while industrial and defense-focused domains may achieve robust Stage~2 or Stage~3 deployments within a few years, wide-scale commercial adoption in consumer-centric domains (particularly those involving dense urban operations or high-severity trucking) remains significantly farther out. Whether these long horizons can be shortened depends on breakthroughs in HPC, reliability methods (improving the \(\beta\) factor in reliability growth models), or regulatory/infrastructure changes. As in all models, real-world constraints and innovation may alter these estimates, but this table offers a structured baseline for understanding the relative timelines across AV categories.

\section{Conclusion}
\label{sec:conclusion}

This paper has presented a \emph{unified} mathematical and system-level approach to estimating autonomous vehicle (AV) timelines by integrating \textbf{computational complexity}, \textbf{reliability growth modeling}, and \textbf{operational design domain (ODD) considerations}. We showed how limits on high-performance computing (HPC), NP-hard multi-agent path planning, safety demonstration constraints, and production/regulatory delays can combine to yield a multi-decade horizon for \emph{fully} universal Level~5 autonomy. While certain niche or restricted ODD applications (e.g., industrial/mining, military/defense) may mature more quickly, broad consumer-focused and high-severity domains (consumer auto, robo-taxis, highway trucking) appear to face significantly longer development pathways under realistic assumptions.

\paragraph{Future Research and Refinements.}
Although we have endeavored to quantify critical parameters such as \(\alpha\), \(\beta\), \(\gamma\), \(\delta\), and \(\chi\) through a mix of theoretical models and practical heuristics, there remains ample room for improvement. Potential avenues of exploration include:
\begin{itemize}
    \item \textbf{Empirical Fitting of Constants:} Gathering more comprehensive real-world data from ongoing pilots and partial deployments to refine \(\alpha\) and \(\beta\) in Crow-AMSAA, or better calibrate \(\chi\) for true HPC reductions.
    \item \textbf{ODD Complexity Estimation:} Creating more detailed and validated methods to assign \(\gamma\) scores across varied urban, suburban, off-road, and highway scenarios; developing domain-specific metrics to capture edge-case density.
    \item \textbf{Parallel-Testing Approaches:} Investigating better ways to conduct reliability growth in tandem with HPC ramp-up, possibly decreasing \(\max\bigl(T_{\mathrm{comp}}, T_{\mathrm{rel}}\bigr)\) in the overall timeline.
    \item \textbf{Regulatory and Infrastructure Synergies:} Exploring how dedicated lanes, sensor-equipped roadways (V2X), or streamlined certification processes might shorten both HPC and reliability demonstration timelines.
    \item \textbf{Improved Simulation-Based Coverage:} Employing next-generation simulation or “digital twin” platforms to accelerate test-mile accumulation, thus increasing effective \(M\) and shrinking long tail corner cases.
\end{itemize}

By incorporating these future directions, the models outlined in this paper could become more precise and better aligned with real-world deployment outcomes. The fundamental framework, however, remains relevant for gauging how AV timelines are shaped by \emph{both} algorithmic/computational limits and statistical reliability targets—an interplay that no single simplified metric can fully capture.

\paragraph{Closing Remarks.}
Even if certain projections here strike the reader as surprisingly distant, they offer a structured baseline against which ongoing developments can be measured. As new data emerges, the parameters (\(\alpha\), \(\beta\), \(\gamma\), \(\chi\), etc.) can be updated, enabling a more precise picture of how quickly (or slowly) different AV categories can move from pilot projects and limited ODDs toward widespread, driverless commercial service. In this sense, the present work is best viewed not as a definitive timeline, but rather as a \emph{scalable model} for tracking the interplay of complexity, reliability, and technology growth in autonomous vehicle development.

\section*{Disclaimer}
These projections rely on current data and theoretical models. Actual timelines may shift with unforeseen breakthroughs (e.g., quantum computing, novel AI paradigms), policy changes, business decisions, procurement issues, or public acceptance factors. Furthermore, the author used a variety of generative AI tools to help identify certain mathematical constructs and assist in document formatting. Nonetheless, the mathematical evidence presented here strongly suggests just how far out autonomy at scale may be.

\clearpage 
\appendix

\section*{Appendix A: AV Timelines Calculation Data}
\label{sec:appendixA}
\addcontentsline{toc}{section}{Appendix A: AV Timelines Calculation Data}
\begin{figure}[!htbp]
    \centering
    \includegraphics[width=1\linewidth]{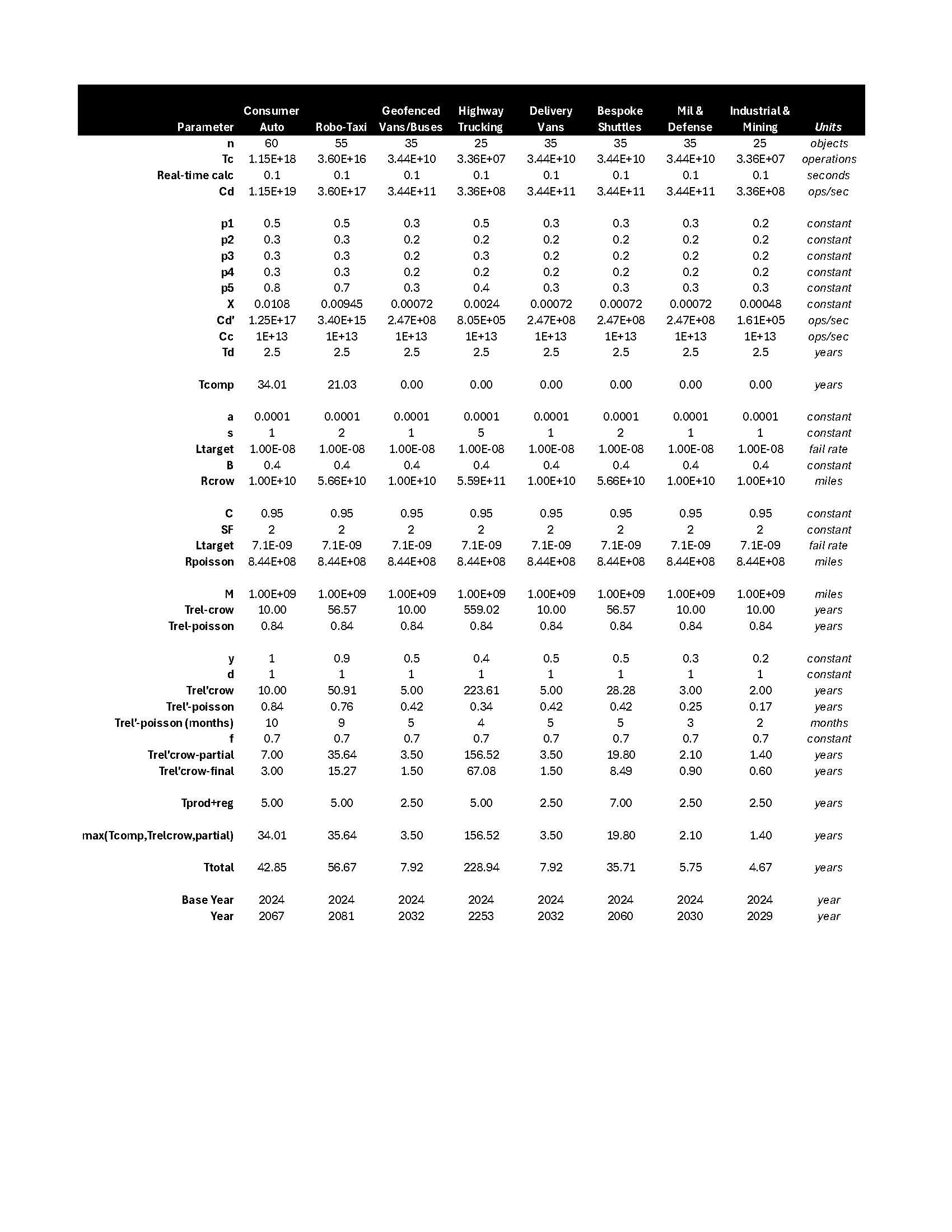}
\end{figure}

\end{document}